\pdfoutput=1 
\documentclass[ 
 aps,%
 twocolumn,%
 reprint,%
 prb,%
 groupedaddress,%
 superscriptaddress,%
 showpacs,%
 lengthcheck,%
 amsfonts,%
 floatfix,%
]{revtex4-1} 

\RequirePackage{amsmath,amssymb,bm}
\RequirePackage{graphicx}
\RequirePackage[]{subfigure}
\RequirePackage{hyperref}
\hypersetup{%
  breaklinks = {true},
  citecolor = {blue},
  colorlinks = {true},
  linkcolor = {red},
  pdfauthor = {\textcopyright\ F\'{a}bio Hip\'{o}lito },
  pdfcreator = {\LaTeX\ and \flqq hyperref\frqq},
  pdffitwindow = {true},
  pdfmenubar = {true},
  pdfpagelayout = {SinglePage},
  pdfstartview = {Fit},
  pdftoolbar = {true},
  plainpages = {false},
}

\allowdisplaybreaks

\newcommand{\ie}{\emph{i.e.}~}
\newcommand{\eg}{\emph{e.g.}~}

\newcommand{\hatH}{\hat{H}}

\newcommand{\Fref}[1]{Fig.~\ref{#1}}
\newcommand{\Fsref}[1]{Figs.~\ref{#1}}
\newcommand{\Eqref}[1]{Eq.~\ref{#1}}
\newcommand{\Eqsref}[1]{Eqs.~\ref{#1}}
\newcommand{\bAcal}{\boldsymbol{\mathcal{A}}}
\newcommand{\bEcal}{\boldsymbol{\mathcal{E}}}
\newcommand{\bNabla}{\boldsymbol{\nabla}}
\newcommand{\bj}{\boldsymbol{\mathit{j}}}
\newcommand{\bG}{\boldsymbol{\Gamma}}

\newcommand{\teq}{{\,=\,}}
\newcommand{\tneq}{{\,\neq\,}}
\newcommand{\tequiv}{{\,\equiv\,}}
\newcommand{\tsim}{{\,\sim\,}}

\DeclareMathOperator{\tr}{tr}

\usepackage[usenames,dvipsnames]{xcolor}

\RequirePackage[normalem]{ulem} 

\graphicspath{ {./Figs/} } 

\begin{document}

\title[Optical THG in black phosphorus]{Optical third harmonic generation in 
black phosphorus}

\author{F. Hipolito}

\email{fh@nano.aau.dk}

\affiliation{Department of Physics and Nanotechnology,
Aalborg University, DK-9220 Aalborg {\O}st, Denmark}

\author{T. G. Pedersen}
\email{tgp@nano.aau.dk}
\affiliation{Department of Physics and Nanotechnology,
Aalborg University, DK-9220 Aalborg {\O}st, Denmark}
\affiliation{Center for Nanostructured Graphene (CNG), 
DK-9220 Aalborg {\O}st, Denmark}

\begin{abstract}
We present a calculation of Third Harmonic Generation (THG) for two-band 
systems using the length gauge that avoids unphysical divergences otherwise 
present in the evaluation of the third order current density response.
The calculation is applied to bulk and monolayer black Phosphorus (bP) using a 
non-orthogonal tight-binding model.
Results show that the low energy response is dominated by mixed inter-intraband 
processes and estimates of the magnitude of THG susceptibility are comparable 
to recent experimental reports for bulk bP samples.
\end{abstract}



\maketitle

\section{Introduction}

Nonlinear light-matter interactions provide a vast field of processes with 
many applications \cite{Shen2002, Boyd2008}, particularly at energies 
comparable to the near IR and visible radiation.
Applications include four wave mixing \cite{Gu2012, Hendry2010}, efficient 
lasing \cite{Wu2015}, harmonic generation, more specifically THG 
\cite{Hong2013, Kumar2013} and Second Harmonic Generation (SHG) in 
non-centrosymmetric crystals, such as transition metal dichalcogenides (TMDs) 
\cite{Li2013a, Zeng2013, Janisch2014, Wang2015, Yin2014, Clark2014} and 
hexagonal Boron Nitride (hBN) \cite{Li2013a}.
Recent advances in atomically thin materials, such as graphene, TMDs and others 
have sparked interest in 2D opto-electronic devices.
The isolation of mono- and few-layer crystals of bP provides new 2D materials 
with remarkable electronic properties, including thickness dependent gap and 
strong in-plane anisotropy.
On its own, the thickness dependent gap of bP \cite{Low2014a, Liu2014, Das2014, 
Li2016a} makes it appealing for opto-electronic devices, since 
its optical gap spans a wide range of the spectrum, from infrared $\sim0.3$ eV 
in bulk samples to visible $\sim1.7$ eV in monolayer \cite{Li2016a}.
Moreover, the low energy dispersion of bP exhibits strong anisotropy, leading 
to a large discrepancy in the effective masses of the valence and conduction 
bands along the armchair and zigzag directions.

The low energy dispersion can be accurately captured by anisotropic massive 
Dirac fermion models \cite{Ezawa2014, Pedersen2017}.
In such systems, electrons effectively behave as light massive Dirac fermions 
along the armchair direction and as heavy fermions along the zigzag direction, 
consistent with \textit{ab-initio} results \cite{Low2014a, Tran2014, Tran2015, 
TaghizadehSisakht2015} and experimental ARPES measurements of the band 
structure \cite{Kim2015}.
The manifestations of anisotropy are tightly connected to the lattice symmetry. 
Both bulk and monolayer bP are orthorhombic crystals with inversion center, 
with space groups $D_{2h}^{18}$ \cite{Takao1981} and $D_{2h}^{7}$ 
\cite{Ribeiro-Soares2014}, respectively.
Due to the presence of an inversion center dipole allowed second order 
interactions are blocked \cite{Shen2002, Boyd2008}, making the THG the leading 
order for harmonic generation.
Recent reports have demonstrated that the electronic and transport properties 
of bP can be used for several applications, including field-effect transitors 
\cite{Li2014a, Koenig2014, Buscema2014a, Das2014}.
The electronic properties of bP provide fertile ground for 
opto-electronics devices, such as photodetectors \cite{Buscema2014a, 
Engel2014}, dichroic absorption \cite{Qiao2014a} and nonlinear optics, 
including THG \cite{Rodrigues2016, Youngblood2017, Autere2017} and high 
harmonic generation \cite{Pedersen2017}.
In addition, theoretical studies indicate that the anisotropic characteristics 
of bP can be harnessed and tuned by strain \cite{Rodin2014, Lv2014a, 
Jiang2017}, opening a door for strain sensitive or strain enhanced 
optoelectronic devices based in bP.

In this work we evaluate the current density response of two-band systems using 
the length gauge \cite{Pedersen2017, Aversa1995} and determine the nonlinear 
THG conductivity tensor.
Moreover, we show that the spurious divergences, present in the straightforward 
evaluation of the nonlinear conductivity, $\sigma_{\phi\lambda\beta\alpha}$, of 
the third order current response \cite{Aversa1995} vanish by considering the 
relevant combinations of $\sigma_{\phi\lambda\beta\alpha}$.
We then use these results to compute and characterize the low energy THG in bP.

\section{Theoretical framework}

We are interested in characterizing the interaction of light with the
electronic system of crystals, within the dipole approximation and 
therefore ignoring the position dependence of the electromagnetic field.
In this approximation, the total Hamiltonian reads
\begin{equation}
\hat{\mathcal{H}} = \hat{\mathcal{H}}_0 +\hat V(t)\,,\, \quad \hat V(t) = e \, 
\hat{\mathbf{r}} \cdot \mathbf{E}(t)\,,
\end{equation}
where $\hat{\mathcal{H}}_0$ defines the unperturbed Hamiltonian for the 
crystal, $\hat V(t)$ contains the time dependent field and $e>0$ is the 
elementary charge.
In addition, the electromagnetic field is monochromatic and linearly polarized
\begin{equation}
\label{eq:E}
 \mathbf{E}(t) = \sum_{\alpha=x,y,z}  \big[ E_{\omega}^\alpha e^{-i\omega t}
 +E_{-\omega}^\alpha e^{ i\omega t } \big] 
 \, \mathbf{e}_\alpha /2 \, ,
\end{equation}
propagating along the $z$-axis, normal to the crystal plane.
The polarization plane defined by the angle relative to the $x$-axis, such that 
$E_{\omega}^\alpha \tequiv E_{\omega}^0 ( \cos\theta, \sin\theta, 0 )$.
The diagonalization of the unperturbed periodic Hamiltonian defines the crystal 
band dispersions $\epsilon_m(\mathbf{k})$ and respective eigenstates, 
$|m,\mathbf{k}\rangle$, which serve as the basis for the calculation of the 
linear and nonlinear response.
The calculation of the response is based on the time dependent density 
operator, $\hat\rho (t) \tequiv \sum_{mn} \rho_{mn} |m\rangle\langle n|$, that 
obeys the quantum Liouville equation
$i\hbar \, \partial \hat \rho / \partial t \teq \big[ \hatH, \hat\rho \big ]$, 
which lends itself to a perturbative expansion.
In this manuscript, we do not consider electron-electron interaction, \eg 
excitonic effects and therefore the many-body effects arise from the 
Fermi-Dirac statistics only.

\subsection{\texorpdfstring{$\pi$}{}-electron tight-binding}
\label{sec:TB}
\begin{figure*}
\includegraphics[width=1.00\linewidth]{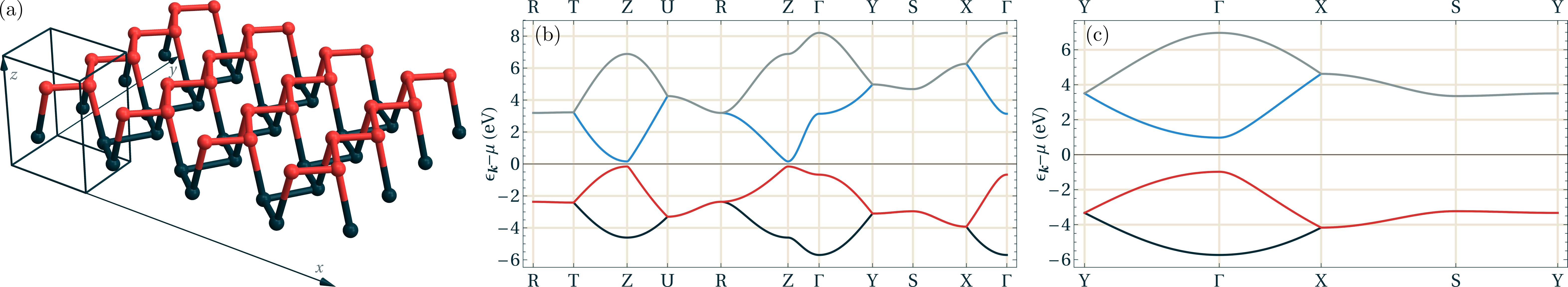}
\caption{\label{fig:1}%
Monolayer lattice for bP (a) and the energy dispersion along high symmetry 
paths for bulk (b) and monolayer (c).}
\end{figure*}

To characterize the low energy properties of bP, we consider a non-orthogonal 
Tight-Binding (TB) model with a $p_z$ orbital per atom in the unit cell. The 
Fourier transforms of the Hamiltonian and the respective overlap matrix read
\begin{subequations}
\begin{align}
\hat H_{ij}(\mathbf{k}) = \sum_{\alpha\beta,\mathbf{R}}
t_{ij}^{\alpha\beta}(\mathbf{r}_i-\mathbf{r}_j+\mathbf{R})
e^{i\mathbf{k} \cdot (\mathbf{r}_i-\mathbf{r}_j+\mathbf{R})} \, ,
\\
\hat S_{ij}(\mathbf{k}) = \sum_{\alpha\beta,\mathbf{R}}
s_{ij}^{\alpha\beta}(\mathbf{r}_i-\mathbf{r}_j+\mathbf{R})
e^{i\mathbf{k} \cdot (\mathbf{r}_i-\mathbf{r}_j+\mathbf{R})} \, ,
\end{align}
\end{subequations}
where $\mathbf{r}_i$ defines the position of $i^\mathrm{th}$ atom in the unit 
cell centered at $\mathbf{R}$.
Furthermore, we consider that the hopping $(t_{ij}^{\alpha\beta})$ and overlap 
$(s_{ij}^{\alpha\beta})$ integrals between orbitals $\{\alpha,\beta\}$ of 
atoms $\{i,j\}$ exhibit spatial dependence like that of Slater--Koster two 
center integrals \cite{Slater1954}.
The above-mentioned integrals are evaluated with density functional 
tight-binding \cite{Porezag1995, Pedersen2013a, Pedersen2017}, using the 
bulk parameters for bP \cite{Takao1981} with a covalent radius of $2.08$ \AA{}.
The lattice is depicted in \Fref{fig:1}a, where the lattice parameters read 
$a_1\teq 4.376$, $a_2\teq 3.314$ and $a_3\teq 5.209$ \AA{} and the respective 
atom positions read
\begin{subequations}
\begin{align}
   \mathbf{r}_1 &= (           -d,   0,           -h ) \, ,
\\ \mathbf{r}_2 &= ( \phantom{-}d,   0, \phantom{-}h ) \, ,
\\ \mathbf{r}_3 &= ( a_1/2 +d, a_2/2, \phantom{-}h ) \, ,
\\ \mathbf{r}_4 &= ( a_1/2 -d, a_2/2,           -h ) \, ,
\end{align}
\end{subequations}
with $d \teq 0.3525$ and $h \teq 1.065$ \AA{} \cite{Takao1981}.
This parametrization leads to energy dispersion consistent with 
\textit{ab-initio} results \cite{Rodin2014,Low2014a,Li2016a} for monolayer, but 
overestimates the bulk gap.
For bulk, the gap \cite{Li2016a} can be recovered by rescaling the coupling 
between different layers with a factor of $\sim0.54$, or conversely by 
stretching the layer separation by $\sim9\%$.
The latter was used to generate all results computed in this work.
Note that we consider normal incidence and as a result, the external field 
couples solely with the in-plane motion of the electrons via the in-plane 
components of the position operator which are not affected by the stretching of 
layer separation.
In \Fsref{fig:1}b and~\ref{fig:1}c, we show the band structures along the 
relevant high symmetry paths for bulk and monolayer bP, respectively. In both 
systems, the TB dispersion is consistent with previous \textit{ab-initio} 
results \cite{Low2014a}.
%
%

Lattice symmetry plays an important role in linear and nonlinear processes as 
it reduces the number of independent and finite tensors elements.
For both bulk and monolayer bP, the optical conductivity is limited to the 
diagonal components $\sigma_{\alpha\alpha}^{(1)}$ \cite{Haussuhl2007}. At third 
order, symmetry reduces the number of independent tensor components to nine 
\cite{Yang1995}, and restricting the external electromagnetic field to normal 
incidence further reduces the number of effective tensor components to four, 
namely 
$\sigma_{11}\tequiv \sigma_{xxxx}$,
$\sigma_{18}\tequiv \sigma_{xxyy} +\sigma_{xyxy} +\sigma_{xyyx}$,
$\sigma_{29}\tequiv \sigma_{yyxx} +\sigma_{yxyx} +\sigma_{yxxy}$ and 
$\sigma_{22}\tequiv \sigma_{yyyy}$.
The combinations of the non-diagonal tensor elements, $\sigma_{18}$ and 
$\sigma_{29}$, will be addressed in detail below, where it is shown that these 
play a crucial role in the calculation of the THG conductivity/susceptibility, 
as these combinations ensure that all non-physical divergences vanish.

\subsection{Perturbative response of two-band systems}

Here, we review the current density response to an external electromagnetic 
field for two-band systems using a perturbative expansion of the time 
dependent density matrix, $\hat\rho(t)$ in the length gauge \cite{Aversa1995, 
Pedersen2015, Hipolito2016} and the single particle velocity operator 
$\hat{\mathbf{v}} = \dot{\hat{\mathbf{r}}} \equiv \hbar^{-1}\bNabla_\mathbf{k} 
\hat{H}$.
The current density for an electronic system with spin degeneracy $g=2$ and 
volume $\Omega$ reads 
$\mathbf{J} = -e g \tr \lbrace \hat{\mathbf{v}} \hat{\rho} \rbrace / \Omega $. 
Upon explicit evaluation of the trace, the current density becomes
\begin{equation}
\label{eq:j:2b}
\mathbf{J} = 
-eg \sum_{\mathbf{k}} \big[
 ( \mathbf{v}_{cc} -\mathbf{v}_{vv} )n/2
+\mathbf{v}_{vc} p +\mathbf{v}_{cv} p^*  \big] / \Omega \, ,
\end{equation}
where we define the population difference $n \equiv \rho_{vv}(t) -\rho_{cc}(t)$ 
and the coherence $p \equiv \rho_{cv}(t)$.
In addition, we made use of the invariance of the trace of the density matrix, 
\ie $\rho_{vv}(t)+\rho_{cc}(t) \teq 1$, together with the fact that the 
integral of the velocity operator over the Brillouin Zone (BZ) vanishes.
The quantum Liouville equation reduces to two dynamical equations for $p$ and 
$n$, namely
\begin{subequations}
\label{eq:QL}
\begin{align}
\label{eq:QL:p}
-i \frac{ \partial p }{ \partial t} + \omega_{cv} p =
-i \mathbf{F}(t) \cdot \big( p \big)_{;\mathbf{k}} 
-  \mathbf{F}(t) \cdot \bAcal_{cv} \, n \, ,
\\
\label{eq:QL:n}
\frac{ \partial n }{ \partial t } =
   \mathbf{F}(t) \cdot \bNabla_\mathbf{k} n
-2i\mathbf{F}(t) \cdot \big( \bAcal_{vc} p -p^* \bAcal_{cv} \big) \, ,
\end{align} 
\end{subequations}
with the condensed notation $\mathbf{F} \equiv -ie \mathbf{E}(t)/(2\hbar)$ and 
$( S_{mn} )_{;\alpha} \equiv \partial S_{mn}/ \partial k_\alpha -i S_{mn} ( 
\mathcal{A}_{mm}^\alpha -\mathcal{A}_{nn}^\alpha ) $ defines the ``generalized 
derivative'' (GD) as in Ref.~\onlinecite{Aversa1995}.
In addition, the matrix elements for the Berry connection read
\begin{equation}
\bAcal_{mn} \equiv \frac{ i }{ \Omega } \int_{ \Omega} \mathrm{d}\mathbf{r} \,
u_{m\mathbf{k}}^*( \mathbf{r} ) \bNabla_\mathbf{k}
u_{n\mathbf{k}}  ( \mathbf{r} )
\, ,
\end{equation}
where $u_{m\mathbf{k}}$ are cell-periodic functions \cite{Hipolito2016}.
The dynamical equations are solved by iteration, generating solutions in 
the form of power series in the external electric field. The iterative
process starts with initial conditions defined by the equilibrium density 
matrix for a cold insulator, \ie absence of coherence $p^{(0)}(t) \teq 0$ and 
fully occupied valence band $n^{(0)}(t) = 1$.
The process is straightforward and has been discussed in detail in 
Refs.~\onlinecite{Aversa1995, Hipolito2016}, hence we display only results for 
the first and third order iterations.
At linear order the difference in the populations is identically zero, 
$n^{(\alpha)} (t) \teq 0$, and the coherence read $p^{(\alpha)} (t) \teq  
p_{\omega}^{(\alpha)} \exp[-i\bar\omega t] +p_{-\omega}^{(\alpha)} \exp[ 
i\bar\omega^* t]$, with Fourier coefficients
\begin{equation}
p_{\omega}^{(\alpha)} =
F_{\omega}^\alpha \mathcal{A}_{cv}^\alpha /( \bar\omega -\omega_{cv} )
\; ,
\end{equation}
where we introduce the complex frequency $\bar\omega\equiv\omega+i\eta$.
The introduction of positive infinitesimal frequency $\eta\equiv 0^+$ in the 
external field ensures the adiabatic switching-on of the interaction 
\cite{Fetter1971}.
At third order, the interaction with an external monochromatic electromagnetic 
field generates two contributions with different fundamental frequencies 
$\lbrace 3\omega, \omega \rbrace$.
The former contributes to the THG and the latter introduces the intensity 
dependent correction to refractive index \cite{Boyd2008, Shen2002}.
The total third order $p(t)$ and $n(t)$ can be cast as
\begin{subequations}
\begin{align}
p^{(\lambda\beta\alpha)} (t) &=  
 p_{ 3\omega}^{(\lambda\beta\alpha)} e^{-3i\bar\omega   t }
+p_{-3\omega}^{(\lambda\beta\alpha)} e^{ 3i\bar\omega^* t }
\nonumber \\&
+p_{  \omega}^{(\lambda\beta\alpha)} e^{- i\bar\omega   t }
+p_{- \omega}^{(\lambda\beta\alpha)} e^{  i\bar\omega^* t } \, ,
\\
n^{(\lambda\beta\alpha)} (t) &=  
 n_{ 3\omega}^{(\lambda\beta\alpha)} e^{-3i\bar\omega   t }
+n_{-3\omega}^{(\lambda\beta\alpha)} e^{ 3i\bar\omega^* t }
\nonumber \\&
+n_{  \omega}^{(\lambda\beta\alpha)} e^{- i\bar\omega   t }
+n_{- \omega}^{(\lambda\beta\alpha)} e^{  i\bar\omega^* t } \, .
\end{align}
\end{subequations}
The relevant THG coherence reads
\begin{subequations}
 \begin{align}
p_{3\omega}^{(\lambda\beta\alpha)} &=
-\frac{ \hbar^3 F_{\omega}^\lambda F_{\omega}^\beta F_{\omega}^\alpha }{ 
3\hbar\bar{\omega} -\epsilon } \Bigg[ 
\frac{\mathcal{A}_{cv}^\lambda }{ 2\hbar\omega } \bigg(
  \frac{ \mathcal{A}_{vc}^\beta  \mathcal{A}_{cv}^\alpha 
  }{ \epsilon -\hbar\omega }
 -\frac{ \mathcal{A}_{vc}^\alpha \mathcal{A}_{cv}^\beta
  }{ \epsilon +\hbar\omega }
\bigg)
\nonumber \\ &+
\bigg( \frac{1}{ 2\hbar\bar{\omega} -\epsilon }
\bigg( \frac{ \mathcal{A}_{cv}^\alpha }{ \hbar\bar{\omega} -\epsilon } 
\bigg)_{;\beta} \bigg)_{;\lambda}\Bigg] \, ,
\end{align}
where we introduce the shorthand notation $\epsilon \tequiv \hbar\omega_{cv}$.
It is important to highlight the presence of a $1/\omega$ divergence in the 
purely interband contribution. This divergence is shown to be spurious in two 
steps, first by isolating the divergent terms by means of partial fraction 
decomposition and then by considering the physical observable, rather than the 
individual components of the density matrix.
With regards to the first step, the coherence becomes
 \begin{align}
\label{eq:p3}
p_{3\omega}^{(\lambda\beta\alpha)} &=
-\frac{ \hbar^3 F_{\omega}^\lambda F_{\omega}^\beta F_{\omega}^\alpha }{ 
3\hbar\bar{\omega} -\epsilon } \Bigg[ 
\frac{\mathcal{A}_{cv}^\lambda }{ 2\epsilon } \bigg[
\frac{ \mathcal{A}_{vc}^\beta \mathcal{A}_{cv}^\alpha 
-\mathcal{A}_{vc}^\alpha \mathcal{A}_{cv}^\beta }{ \hbar\bar{\omega} }
\nonumber \\&+
\frac{ \mathcal{A}_{vc}^\beta \mathcal{A}_{cv}^\alpha ( \epsilon 
+\hbar\bar{\omega} )
+\mathcal{A}_{vc}^\alpha \mathcal{A}_{cv}^\beta ( \epsilon -\hbar\bar{\omega} )
}{ \epsilon^2 -\hbar^2{\bar{\omega}}^2 }
\bigg] 
\nonumber \\ &+
\bigg( \frac{1}{ 2\hbar\bar{\omega} -\epsilon }
\bigg( \frac{ \mathcal{A}_{cv}^\alpha }{ \hbar\bar{\omega} -\epsilon } 
\bigg)_{;\beta} \bigg)_{;\lambda}\Bigg] \, .
\end{align}
\end{subequations}
In the context of light-matter interaction, the current density 
\Eqref{eq:j:2b} (or the respective polarization density) represents the 
physical observable, more specifically the THG Fourier components read
\begin{align*}
j_\phi(3\omega) &= 
-\frac{eg}{\Omega}
\sum_\mathbf{k} \sum_{mn} 
v_{nm}^\phi \sum_{\lambda\beta\alpha} \rho_{mn}^{(\lambda\beta\alpha)}
\\&
= \sum_{\lambda\beta\alpha} \sigma_{\phi\lambda\beta\alpha}(3\omega)
E_{\omega}^\lambda E_{\omega}^\beta E_{\omega}^\alpha \, ,
\end{align*}
which in turn defines the rank-4 tensor.
Moreover, the physically relevant elements of a general rank-4 tensor in three 
dimensions can be grouped into thirty effective tensors according to the 
dependence on the external field \cite{Yang1995}. This can be summarized in 
three classes according to the combinations of indices 2, 3 and 4:
\begin{itemize}
 \item $\sigma_{\phi\alpha\alpha\alpha}$: $9$ individual components, $3$ 
 diagonal ($\alpha \teq \phi$) and $6$ with three repeated entries 
 ($\alpha\tneq\phi$); 
 \item $\sigma_{\phi\beta\alpha\alpha}+\sigma_{\phi\alpha\beta\alpha} 
 +\sigma_{\phi\alpha\alpha\beta}$: $3\times8\teq18$ combinations with two 
 repeated entries ($\alpha$ appears twice) in tensor indices 2, 3 and 4;
 \item $\sigma_{\phi\lambda\beta\alpha}+\sigma_{\phi\lambda\alpha\beta}
       +\sigma_{\phi\alpha\lambda\beta}+\sigma_{\phi\alpha\beta\lambda}
       +\sigma_{\phi\beta\alpha\lambda}+\sigma_{\phi\beta\lambda\alpha}$: $3$ 
 combinations with no repeating entries in tensor indices 2, 3 and 4.
\end{itemize}
By considering these combinations, it becomes clear that the divergence in the 
coherence (\Eqref{eq:p3}) is spurious, as the $1/\omega$ terms add up to zero.
Therefore, the divergent term can be removed from the original definition, and 
thus define the divergence free effective density matrix $\langle \rho 
\rangle$, \eg in tensors with two repeating entries $\langle 
\rho^{(\beta\alpha\alpha)} \rangle = \rho^{(\beta\alpha\alpha)}
+\rho^{(\alpha\beta\alpha)} +\rho^{(\alpha\alpha\beta)}$.

With regards to $n_{3\omega}^{(\lambda\beta\alpha)}$, the dynamical equation 
leads to a rather lengthy and cumbersome expression that contains $1/\omega$ 
divergences.
As in the case of $p_{3\omega}^{(\lambda\beta\alpha)}$, these divergences are 
shown to vanish for the physically relevant combinations of the 
$\sigma_{\phi\lambda\beta\alpha}$.
The process of extricating the spurious terms is made simpler by expanding 
the numerator in a power series of the photon energy, which naturally isolates 
the divergent terms
\begin{align}
\label{eq:n3}
n_{3\omega}^{(\lambda\beta\alpha)} &=
F_{\omega}^\lambda F_{\omega}^\beta F_{\omega}^\alpha
\frac{ i\hbar^{3}\,\sum_{j=0}^5 ( \hbar\bar\omega )^{j-1} 
n_j^{\lambda\beta\alpha} }{ 
3\epsilon^2 
( \hbar^2\bar\omega^2 -\epsilon^2 )^2 ( 4\hbar^2\bar\omega^2 -\epsilon^2 ) }
\, ,
\end{align}
where coefficients $n_j^{\lambda\beta\alpha}$ are frequency independent and 
retain the tensorial nature of $n_{3\omega}^{(\lambda\beta\alpha)}$.
The respective elements are expressed in terms of the gauge invariant 
GD \cite{Aversa1995}, 
\begin{widetext}
\begin{subequations}
\label{eq:n3:nums}
\begin{align}
 n_0^{\lambda\beta\alpha} &= 
2\epsilon^5 \Big[ 
 (\mathcal{A}_{vc}^\lambda \mathcal{A}_{cv}^\alpha 
 +\mathcal{A}_{vc}^\alpha  \mathcal{A}_{cv}^\lambda )
\, \partial \epsilon / \partial k_\beta 
-(\mathcal{A}_{vc}^\beta   \mathcal{A}_{cv}^\alpha 
 +\mathcal{A}_{vc}^\alpha  \mathcal{A}_{cv}^\beta )
\, \partial \epsilon / \partial k_\lambda
 \Big]
+\epsilon^6 \Big[
  \big( \mathcal{A}_{vc}^\beta  \mathcal{A}_{cv}^\alpha 
 +      \mathcal{A}_{vc}^\alpha \mathcal{A}_{cv}^\beta  \big)_{;\lambda} 
\nonumber\\ &
 -2\mathcal{A}_{vc}^\lambda ( \mathcal{A}_{cv}^{\alpha} )_{;\beta}
 -2( \mathcal{A}_{vc}^{\alpha} )_{;\beta}   \mathcal{A}_{cv}^\lambda \Big] \; ,
\\
n_1^{\lambda\beta\alpha} &=
-\epsilon^4 \Big[ 
 3( \mathcal{A}_{vc}^\alpha  \mathcal{A}_{cv}^\beta 
   -\mathcal{A}_{vc}^\beta   \mathcal{A}_{cv}^\alpha )
 \, \partial \epsilon / \partial k_\lambda
+8( \mathcal{A}_{vc}^\alpha  \mathcal{A}_{cv}^\lambda
   -\mathcal{A}_{vc}^\lambda \mathcal{A}_{cv}^\alpha )
 \, \partial \epsilon / \partial k_\beta
\Big]
+\epsilon^5 \Big[
6\big[
 \big( \mathcal{A}_{vc}^\alpha \big)_{;\beta} \mathcal{A}_{cv}^\lambda 
-\mathcal{A}_{vc}^\lambda \big( \mathcal{A}_{cv}^\alpha \big)_{;\beta} \big]
\nonumber \\ &
 +\big( \mathcal{A}_{vc}^\alpha  \mathcal{A}_{cv}^\beta
 -      \mathcal{A}_{vc}^\beta   \mathcal{A}_{cv}^\alpha \big)_{;\lambda}
\Big] \; ,
\\
n_2^{\lambda\beta\alpha} &= 
\epsilon^3 \Big[
 10( \mathcal{A}_{vc}^\lambda \mathcal{A}_{cv}^\alpha 
    +\mathcal{A}_{vc}^\alpha  \mathcal{A}_{cv}^\lambda )
 \, \partial \epsilon / \partial k_\beta
+8 ( \mathcal{A}_{vc}^\beta   \mathcal{A}_{cv}^\alpha
    +\mathcal{A}_{vc}^\alpha  \mathcal{A}_{cv}^\beta  )
 \, \partial \epsilon / \partial k_\lambda
\Big]
-\epsilon^4 \Big[
2\big[ 
 \big( \mathcal{A}_{vc}^\alpha \big)_{;\beta} \mathcal{A}_{cv}^\lambda
+\mathcal{A}_{vc}^\lambda \big( \mathcal{A}_{cv}^\alpha \big)_{;\beta} \big]
\nonumber \\ &
+5\big( \mathcal{A}_{vc}^{\beta}  \mathcal{A}_{cv}^\alpha 
 +\mathcal{A}_{vc}^{\alpha} \mathcal{A}_{cv}^\beta \big)_{;\lambda}
\Big] \; ,
\\
n_3^{\lambda\beta\alpha} &= 
\epsilon^2 \Big[
  4( \mathcal{A}_{vc}^\lambda \mathcal{A}_{cv}^\alpha 
    -\mathcal{A}_{vc}^\alpha  \mathcal{A}_{cv}^\lambda )
 \, \partial \epsilon / \partial k_\beta
+13( \mathcal{A}_{vc}^\alpha  \mathcal{A}_{cv}^\beta
    -\mathcal{A}_{vc}^\beta   \mathcal{A}_{cv}^\alpha )
 \, \partial \epsilon / \partial k_\lambda
\Big]
+\epsilon^3 \Big[
 6\big[ 
 \mathcal{A}_{vc}^\lambda \big( \mathcal{A}_{cv}^\alpha \big)_{;\beta}
-\big( \mathcal{A}_{vc}^\alpha \big)_{;\beta} \mathcal{A}_{cv}^\lambda \big]
\nonumber \\ &
+5\big( \mathcal{A}_{vc}^{\beta}  \mathcal{A}_{cv}^\alpha 
 -\mathcal{A}_{vc}^{\alpha} \mathcal{A}_{cv}^\beta \big)_{;\lambda}
\Big] \; ,
\\
n_4^{\lambda\beta\alpha} &= 
4 \epsilon^2 \Big[
\big( \mathcal{A}_{vc}^\alpha \big)_{;\beta} \mathcal{A}_{cv}^\lambda 
+\mathcal{A}_{vc}^\lambda \big( \mathcal{A}_{cv}^\alpha \big)_{;\beta}
+\big( \mathcal{A}_{vc}^{\beta}  \mathcal{A}_{cv}^\alpha 
 +\mathcal{A}_{vc}^{\alpha} \mathcal{A}_{cv}^\beta \big)_{ ;\lambda }
\Big] \; ,
\\
n_5^{\lambda\beta\alpha} &= 
4 \epsilon^0 \, \frac{ \partial \epsilon }{ \partial k_\lambda }
( \mathcal{A}_{vc}^{\beta}  \mathcal{A}_{cv}^\alpha 
 -\mathcal{A}_{vc}^{\alpha} \mathcal{A}_{cv}^\beta )
-4 \epsilon^1 \big( \mathcal{A}_{vc}^{\beta}  \mathcal{A}_{cv}^\alpha 
 -\mathcal{A}_{vc}^{\alpha} \mathcal{A}_{cv}^\beta \big)_{ ;\lambda }
\; ,
\end{align}
\end{subequations} 
\end{widetext}
however several terms reduce to regular derivatives, as the Berry 
connection part of the GD vanishes.
Following the procedure outlined above for the coherence, it is 
straightforward to show that the contributions from the effective coefficients 
$\langle n_0^{\lambda\beta\alpha} \rangle$ vanish, thus showing that the 
$1/\omega$ divergence is spurious.
Additional spurious contributions are found in the higher order terms of 
this expansion. Discarding these contributions allows for the simplification of 
several terms, namely $ \langle n_1^{\lambda\beta\alpha} \rangle \equiv 
-\epsilon^2 \langle n_3^{\lambda\beta\alpha} \rangle \equiv 6 \epsilon^5 \big[ 
\big( \mathcal{A}_{vc}^\alpha \big)_{;\beta} \mathcal{A}_{cv}^\lambda 
-\mathcal{A}_{vc}^\lambda \big( \mathcal{A}_{cv}^\alpha \big)_{;\beta} \big]$ 
and $ \langle n_5^{\lambda\beta\alpha}  \rangle\tequiv 0 $.

Based on the regularized expressions for the coherence and population 
difference, we define the THG conductivity as a combination of three terms
$\sigma_{\phi\lambda\beta\alpha}^{(3)} = 
\sigma_{\phi\lambda\beta\alpha}^{(3,A)}+\sigma_{\phi\lambda\beta\alpha}^{(3,B)} 
+\sigma_{\phi\lambda\beta\alpha}^{(3,C)}$
separated according to the nature of the transitions involved in each term.
Contributions arising from purely interband transitions are captured in the 
first term, $A$, whereas the remaining terms concern mixed processes, 
involving \emph{one} or \emph{two} intraband transitions, $B$ and $C$ 
respectively.
The full form of each contribution becomes
\begin{subequations}
\label{eq:s3}
\begin{align}
\label{eq:s3:a}
\frac{ \sigma_{\phi\lambda\beta\alpha}^{(3,A)} (3\omega)
}{ i\, \sigma_3 \mathcal{N}_d } &= 
\hbar^3 \sum_\mathbf{k}
\frac{ v_{vc}^\phi v_{cv}^\lambda }{ 3\hbar\bar{\omega} -\epsilon }
\frac{ v_{vc}^\beta v_{cv}^\alpha +v_{vc}^\alpha v_{cv}^\beta }{
\epsilon^{3}( \hbar^2{\bar{\omega}}^2 -\epsilon^2 ) }
%
%
+(c \leftrightarrow v)
\\
\label{eq:s3:b}
\frac{ \sigma_{\phi\lambda\beta\alpha}^{(3,B)} (3\omega)
}{ i\, \sigma_3 \mathcal{N}_d } &= 
\sum_\mathbf{k}
\frac{ v_{cc}^\phi -v_{vv}^\phi }{ 4\hbar^2\bar\omega^2 -\epsilon^2 }
%
%
\frac{ \sum_{j=0}^5 ( \hbar\bar\omega )^{j-1} n_j^{\lambda\beta\alpha}
}{ 3( \hbar^2\bar\omega^2 -\epsilon^2 )^2 }
\\
\label{eq:s3:c}
\frac{ \sigma_{\phi\lambda\beta\alpha}^{(3,C)} (3\omega)
}{ i\, \sigma_3 \mathcal{N}_d } &= 
\hbar \sum_\mathbf{k}
\bigg( \frac{ v_{vc}^\phi }{ 3\hbar\bar{\omega} -\epsilon } 
\bigg)_{;\lambda}
\frac{ 1 }{ 2\hbar\bar{\omega} -\epsilon }
%
%
\bigg( \frac{ v_{cv}^\alpha / \epsilon }{ \hbar\bar{\omega} 
-\epsilon } \bigg)_{;\beta}
+\nonumber\\ &+(c \leftrightarrow v)
\, ,
\end{align}
\end{subequations}
where the interband position matrix elements are expressed as velocity matrix 
elements via $\mathcal{A}_{mn}^\alpha \teq -i \hbar \, v_{mn}^\alpha /
\epsilon_{mn}$ \cite{Aversa1995,Hipolito2016}. $\mathcal{N}_d$ is a 
normalization constant and $\sigma_3$ sets the scale of the THG conductivity.
Given that the dimensionality of the system under consideration defines the 
dimensions of $\sigma^{(N)}$ and $\chi^{(N)}$, we choose to set the $\sigma_3$ 
and $\mathcal{N}_d$ for 2D systems.
In 2D the THG conductivity scale reads $\sigma_3 = e^4 a_0^2 / (8 \gamma_0^2 
\hbar) = 3.04\times 10^{-25}$ $\mathrm{Sm^2/V^2}$, with $\gamma_0 =  
1\,\mathrm{eV}$, $a_0 = 1$ \AA{}. 
The respective normalization constant $\mathcal{N}_2 \equiv g \gamma_0^2 
\hbar/( a_0^2 A )$, where $A \equiv A_C N_x N_y $ is the total area for $N_x 
N_y$ unit cells with area $A_C$.
For the 3D system, the normalization constant is defined as $\mathcal{N}_3 
\equiv a_3 g \gamma_0^2 \hbar/( a_0^2 N_x N_y N_z V_C ) = \mathcal{N}_2/N_z$, 
with unit cell volume $V_C=a_3A_C$ and $N_z$ unit cells along the 
$z$-direction. The conversion of 3D to 2D nonlinear conductivity is obtained 
through the multiplication by the vertical lattice parameter $a_3$.
Moreover, to improve numerical stability and account for broadening in 
realistic spectra, we keep the adiabatic coupling finite, $\hbar\eta = 0.05$ 
eV, throughout all calculations.
It is worth mentioning that in case of the diagonal tensor elements, the $A$ 
and $B$ contributions reduce to compact closed-form expressions
\begin{subequations}
\label{eq:s3diag}
 \begin{align}
\label{eq:s3diag:a}
\frac{ \sigma_{\phi\phi\phi\phi}^{(3,A)} (3\omega) 
}{ i\, \sigma_3 \mathcal{N}_d } 
= 
\hbar^3  \sum_\mathbf{k}
\frac{ 12 \hbar \bar\omega \, | v_{vc}^\phi |^4 / \epsilon^{3} }{
( 9\hbar^2{\bar{\omega}}^2 -\epsilon^2 )
(  \hbar^2{\bar{\omega}}^2 -\epsilon^2 ) }
\\
\label{eq:s3diag:b}
\frac{ \sigma_{\phi\phi\phi\phi}^{(3,B)} (3\omega) 
}{ i\, \sigma_3 \mathcal{N}_d  } 
= 
\hbar^2 \sum_\mathbf{k}
\frac{ 2( v_{cc}^\phi -v_{vv}^\phi )\epsilon^2 }{ 
( 4\hbar^2\bar\omega^2 -\epsilon^2 )( \hbar^2\bar\omega^2 -\epsilon^2 ) }
\nonumber\\ \times
\bigg[
\frac{ 6\hbar\bar\omega |v_{vc}^\phi|^2 /\epsilon 
}{ \hbar^2\bar\omega^2 -\epsilon^2 }
\frac{\partial\epsilon}{\partial k_\phi} 
+ \bigg( \frac{ v_{vc}^\phi }{ \epsilon } \bigg)_{;\phi}
\frac{ v_{cv}^\phi (2\hbar\bar\omega-\epsilon) }{ \epsilon }
+\nonumber\\+
\frac{ v_{vc}^\phi (2\hbar\bar\omega-\epsilon) }{ \epsilon }
\bigg( \frac{ v_{cv}^\phi }{ \epsilon } \bigg)_{;\phi}
\bigg] \, ,
\end{align}
\end{subequations}
that allow for a more clear understanding of the nature of each process.

Under irradiation by an external electromagnetic field, the linear and 
nonlinear optical conductivities generate currents in the material, which in 
turn radiate an electromagnetic field, $\bEcal(t)$, that includes among other 
contributions the $n^\mathrm{th}$ harmonic field \cite{Boyd2008, Shen2002}.
For a thin sheet in the interface of two media, the currents radiate a flux 
density $I(\omega) =\varepsilon_0 c |\bEcal_\omega|^2/2 = 
\mu_0c|\bj(\omega)|^2/8$ \cite{Hipolito2017, Hipolito2017a}, that can be 
analyzed with a linear polarizer, such that the flux density transmitted 
through the linear polarizer reads 
$ I_\zeta(\omega) = \mu_0c|\bj(\omega)\cdot(\cos\zeta,\sin\zeta,0)|^2/8$.
The latter provides a tool to analyze $n^\mathrm{th}$ harmonic generation as it 
allows to disentangle the contributions from different tensor elements, using 
exclusively optical techniques.
For third order processes in orthorhombic crystals, with the external field 
linearly polarized at an angle $\theta$ with respect to the $x$-axis, the 
intensity of the filtered signals along $x$ ($\zeta=0$) and $y$ ($\zeta=\pi/2$) 
read
\begin{subequations}
\label{eq:I3}
\begin{align}
\label{eq:I3:x}
I_x / I_0 =
  | \bar\sigma_{11} |^2 \cos^6\theta 
+2 \Re[ \bar\sigma_{11} \bar\sigma_{18}^* ] \cos^4\theta \sin^2\theta \, ,
\\
\label{eq:I3:y}
I_y / I_0 =
  | \bar\sigma_{22} |^2 \sin^6\theta
+2 \Re[ \bar\sigma_{22} \bar\sigma_{29}^* ] \cos^2\theta \sin^4\theta \, ,
\end{align}
\end{subequations}
where $I_0 \teq \mu_0 c \sigma_3^2 E_0^6/8$ sets the intensity scale, with
$\bar\sigma_{ij} \equiv \sigma_{ij}/\sigma_3$.
\Eqsref{eq:I3} can be used to probe the magnitudes of effective tensor 
elements and a couple of relative phases from experimental data. Additional 
relative phases can be determined by measuring the so-called parallel and 
perpendicular intensity, \ie analyzer synchronized with the polarization plane 
such that $\zeta=\theta$ and $\zeta=\theta+\pi/2$ for parallel and 
perpendicular intensities.

\section{Results}
We start by addressing the key properties of the energy dispersion of the 
$\pi$-electron tight-binding model for bulk and monolayer.
\Fref{fig:1}b shows the bulk energy dispersion along a high-symmetry path 
in the orthorhombic BZ, with chemical potential $\mu = -5.31$ eV.
It exhibits a direct gap, $E_g =\epsilon_{cv}(\mathbf{k=\mathbf{Z}}) =0.316$ 
eV, at the $\mathbf{Z} = (0,0,\pi/a_3)$ point and the second lowest resonant 
vertical transition is associated with the $\bG$ point has a much larger energy 
separation $\Delta E = 3.80$ eV. 
Therefore, the low energy ($\hbar\omega \simeq 1$ eV) optical response, 
including THG, should depend mostly on transitions associated with the vicinity 
of the $\mathbf{Z}$ point. 
With regards to the monolayer, the energy dispersion is shown in 
\Fref{fig:1}c, with $\mu = -4.75$ eV. It also exhibits a direct gap, 
$E_g = 1.95$ eV found at the BZ center $\bG$.
Moreover, the relative difference to the next resonant vertical transition, 
$\Delta E(\mathbf{k}=\mathbf{S}) = 6.58$ eV is significantly smaller than in 
bulk, where $\mathbf{S}=\pi(a_1^{-1},a_2^{-1},0)$. As discussed below, 
transitions occurring in the vicinity of $\mathbf{S}$ can play a role in THG at 
the energy scale of the gap, \ie $\hbar\omega \sim E_g$.

Regarding the optical properties, we start by considering the optical 
conductivity, evaluated with Eq. 22 of Ref. \onlinecite{Hipolito2016}.
%
\begin{figure}
\centering
\includegraphics[width=1.00\linewidth]{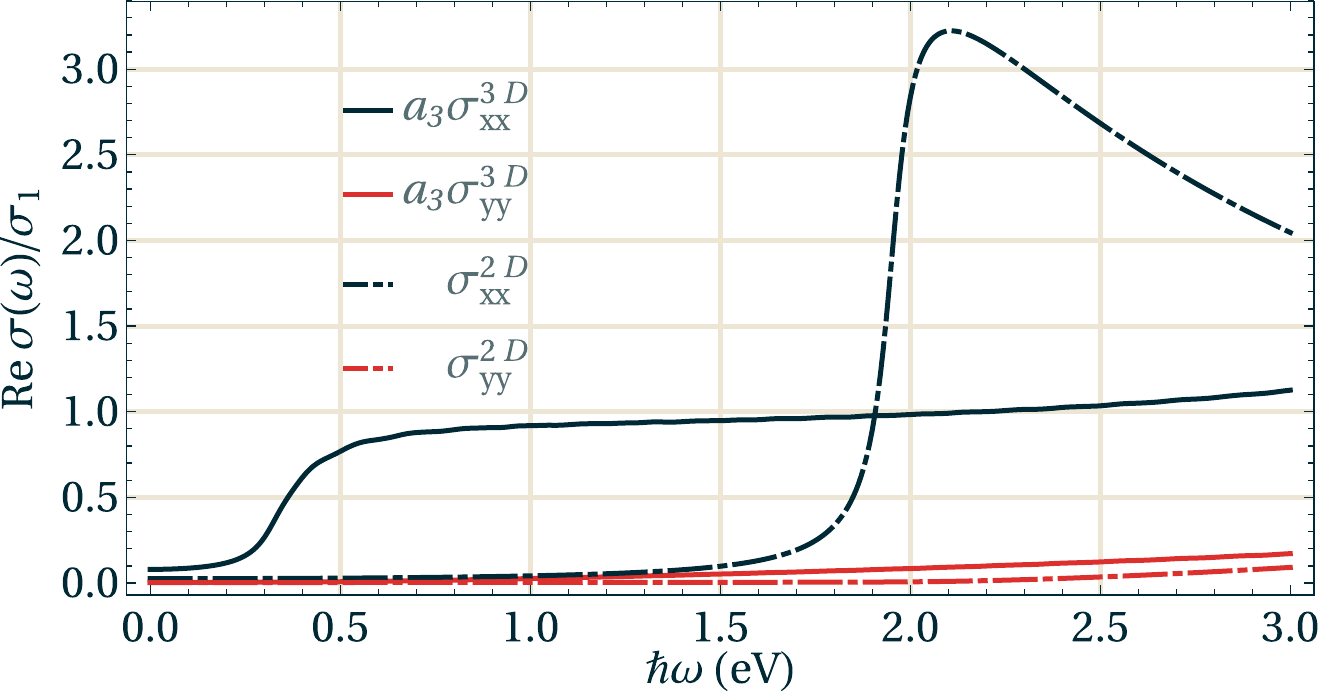}
\caption[Optical conductivity and FH intensity]{\label{fig:2}%
Linear response of bulk and monolayer bP $\hbar\omega = 0.793 \, 
\mathrm{eV} \, \sim 1560 \, \mathrm{nm}$.
The conductivity is plotted in units of 2D conductivity $\sigma_1 = 
e^2/4h $, where the bulk conductivity is converted into 2D conductivity by 
multiplying by the vertical lattice parameter $a_3 = 5.209$ \AA.
%
}
\end{figure}
%
%
In \Fref{fig:2}, we plot the real part of diagonal elements of the conductivity 
tensor, $\sigma_{xx}$ (black) and $\sigma_{yy}$ (red), with solid lines and 
dot-dashed lines representing the bulk and monolayer responses.
The off-diagonal conductivity elements are identically zero, as expected for 
crystals with inversion symmetry.
The lattice anisotropy manifests itself similarly in bulk and monolayer 
systems, where the $|\sigma_{xx}|/|\sigma_{yy}|\sim 20$ ratios at the respective 
band gap threshold, $\hbar\omega \sim E_g$, exhibit the dominant nature of 
$\sigma_{xx}$ at low energy.
In spite of the clearly distinct frequency dependence, results show (upon 
conversion to a 2D conductivity) that the bulk response has a magnitude 
comparable to that of the monolayer and to the quantum of conductance $\sim 
\sigma_1 = e^2/4\hbar$.
The presence of the finite broadening energy, $\hbar\eta = 0.05$ eV, smoothens 
the response at the optical gap and is responsible for the apparently finite 
conductivity at zero frequency in the bulk results \cite{Low2014a, Li2016a}.
The optical conductivity of bulk is in agreement with reports on extinction 
spectra \cite{Zhang2017} and with the dielectric function computed from 
Electron Energy Loss Spectroscopy (EELS) data \cite{Schuster2015}.
Results for monolayer are consistent with previous calculations in the single 
particle approximation \cite{Low2014a, Tran2014, Tran2015}, but show 
limitations of this approximation by not accounting for excitonic resonances 
present of monolayer bP \cite{Wang2015c, Li2016a, Zhang2017}.
%
%
\begin{figure}
\centering
\includegraphics[width=1.00\linewidth]{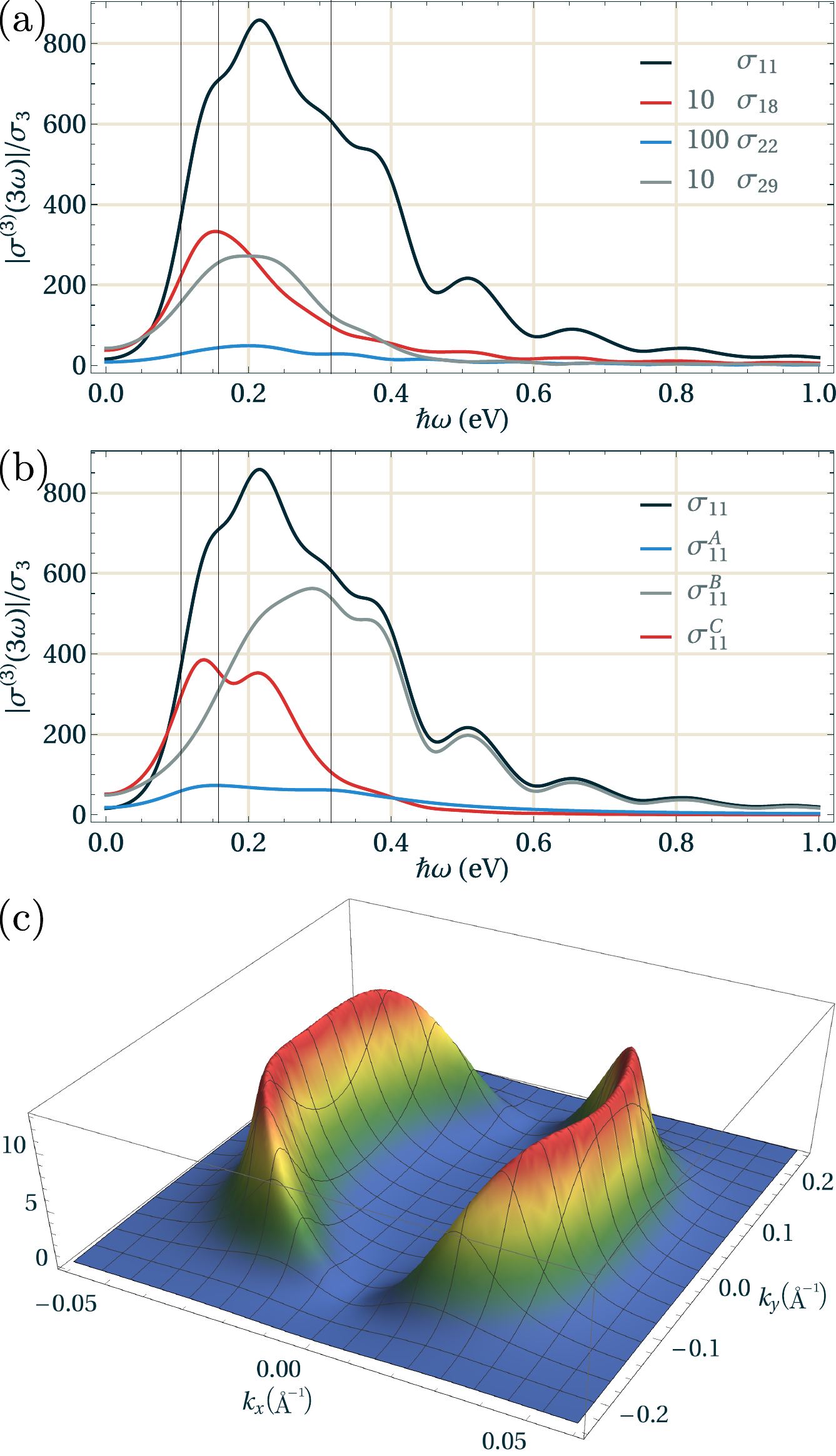}
\caption[THG conductivity for bulk]{\label{fig:3}%
THG of bulk bP in dimensions of 2D nonlinear conductivity $\sigma_3 $.
In (a), we plot the absolute value of the effective tensor components of 
$\sigma^{(3)}(3\omega)$.
Curves for $\sigma_{18}$, $\sigma_{29}$ are scaled by a factor of 10 and 
$\sigma_{22}$ by 100.
Black vertical lines indicate THG resonances, $\hbar\omega =E_g/3,E_g/2,E_g$. 
(b) illustrates the decomposition of the dominant term, $\sigma_{11}$, into the 
components of \Eqsref{eq:s3}.
(c) shows the magnitude (in arbitrary units) integrand 
($\phi\lambda\beta\alpha=xxxx$) of \Eqref{eq:s3:b} near $\mathbf{Z}$ with 
$\hbar\omega =0.5\,\mathrm{eV}$ and $k_z = \pi/a_3$.
}
\end{figure}%
%
%

With respect to THG, \Fref{fig:3}a shows the magnitude of the four effective 
nonlinear conductivity tensors, namely $\sigma_{11}$, $\sigma_{18}$, 
$\sigma_{29}$ and $\sigma_{22}$ as discussed in \S\ref{sec:TB}.
The THG is, similarly to the linear response, highly anisotropic and dominated 
by response along the $x$-axis, \ie $\sigma_{11}$.
To make the remaining effective conductivities visible in \Fref{fig:3}a, we 
amplify $\sigma_{18}$, $\sigma_{29}$ by a factor of 10 and $\sigma_{22}$ by 100.
\Fref{fig:3}b is dedicated to the analysis of the dominant term, $\sigma_{11}$, 
where we compare the magnitude with the individual contributions, as defined in 
\Eqref{eq:s3}.
Results show that the response in the low energy range is dominated by the 
mixed inter-intraband processes. The double intraband process, 
\Eqref{eq:s3:c}, plays an important role at very low energies and decays 
rapidly for higher energies.
On the other hand, the single intraband process, \Eqref{eq:s3:b}, generates the 
overall largest contribution and contains multiple resonances including 
some above the band gap energy.
It is worth noticing that all resonances are blue shifted with respect to the 
band gap resonances, \ie $\hbar\omega=E_g/3,E_g/2,E_g$.
In \Fref{fig:3}c, we plot a map of the absolute value of the integrand present 
in \Eqref{eq:s3:b} in the vicinity of the high symmetry point $\mathbf{Z}$ at 
$\hbar\omega = 0.5$ eV. 
This behavior is common for all integrands independently of the photon energy 
and leads to the blocking of the lowest energy transitions, which in turn 
causes the blue shift of the resonances.
Additionally, it identifies the contributions that generate various features in 
the THG response, such as the peak at $\hbar\omega\sim 0.5$ eV.
The vanishing nature of the integrands of \Eqsref{eq:s3} at the $\mathbf{Z}$ 
point stems from three different sources that individually exhibit this 
behavior.
First, products of the velocity matrix elements, such as $v_{vc}^\beta 
v_{cv}^\alpha$.
Second, difference between diagonal velocity matrix, \eg 
$v_{cc}^\alpha-v_{vv}^\alpha$.
Third, all gradients and GDs present in \Eqsref{eq:s3}.

%
\begin{figure}
\centering
\includegraphics[width=1.00\linewidth]{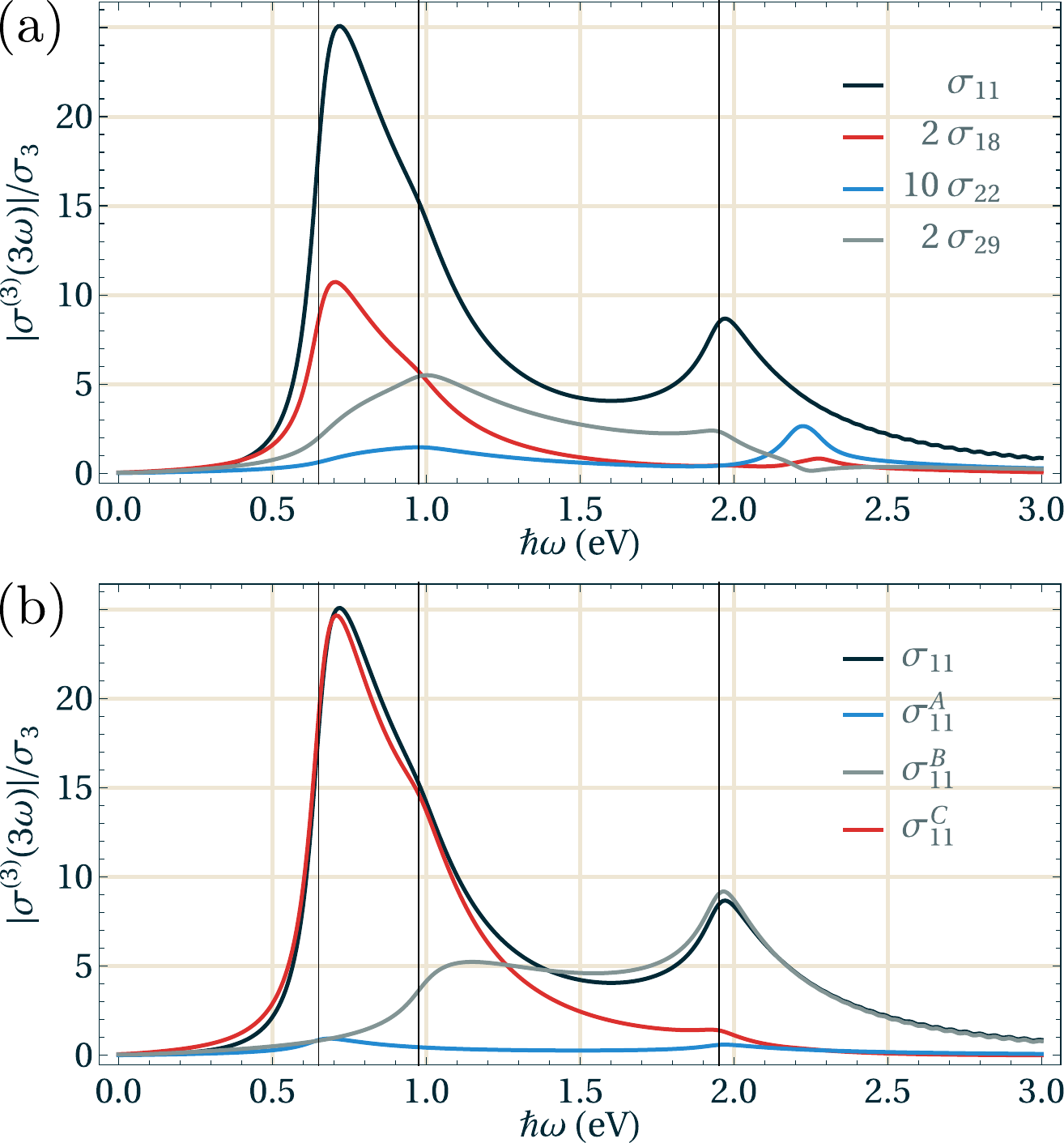}
\caption[THG conductivity for monolayer]{\label{fig:4}%
THG of monolayer bP.
In (a), we plot the absolute value of the effective tensor components of 
$\sigma^{(3)}(3\omega)$.
Curves for $\sigma_{18}$, $\sigma_{29}$ are scaled by a factor of 2 and 
$\sigma_{22}$ by 10.
Black vertical lines indicate THG resonances, $\hbar\omega =E_g/3,E_g/2,E_g$. 
(b) shows the decomposition of the dominant term into the components 
\Eqsref{eq:s3}.
}
\end{figure}%

Turning our attention to the monolayer, \Fref{fig:4}a shows the magnitude 
of the four effective THG conductivities.
The monolayer THG response exhibits several differences with respect to the 
bulk response.
First, all features appear at resonances associated with a large joint 
density of states, including the small resonance slightly above the band gap 
energy, $\hbar\omega =\epsilon_{cv}(\mathbf{k} =\mathbf{S})/3 \sim2.19$ eV.
The presence of the latter shows that the entire BZ contributes to the THG at 
the energy scale of the fundamental resonance $\hbar\omega\sim E_g$.
Second, \Fref{fig:4}b shows that the THG conductivity is dominated by the mixed 
processes but, unlike in the bulk, each term dominates in distinct parts of the 
spectrum with minimal overlap near the resonance $2\hbar\omega\sim E_g$.
The lowest energy response is dominated by the doubly intraband process, 
whereas the response in the vicinity of the gap threshold is controlled by the 
single intraband process.
Moreover, the largest magnitude of the nonlinear conductivity is found at the 
lowest resonance, $3\hbar\omega\sim E_g$.
Last, but not least, the overall scale of the THG conductivity is significantly 
smaller than that of the bulk crystal, \eg the ratio between the maximum THG 
conductivities is $\sim35$.
This can be understood as a consequence of the decay of the nonlinear 
conductivity with the increase of the gap, as in the case of the second order 
response \cite{Hipolito2016}.
Yet, due to the intricate nature of \Eqsref{eq:s3}, it was not possible to 
determine an accurate estimate for the gap dependence of the THG conductivity 
in bP.

\begin{figure}
\centering
\includegraphics[width=1.00\linewidth]{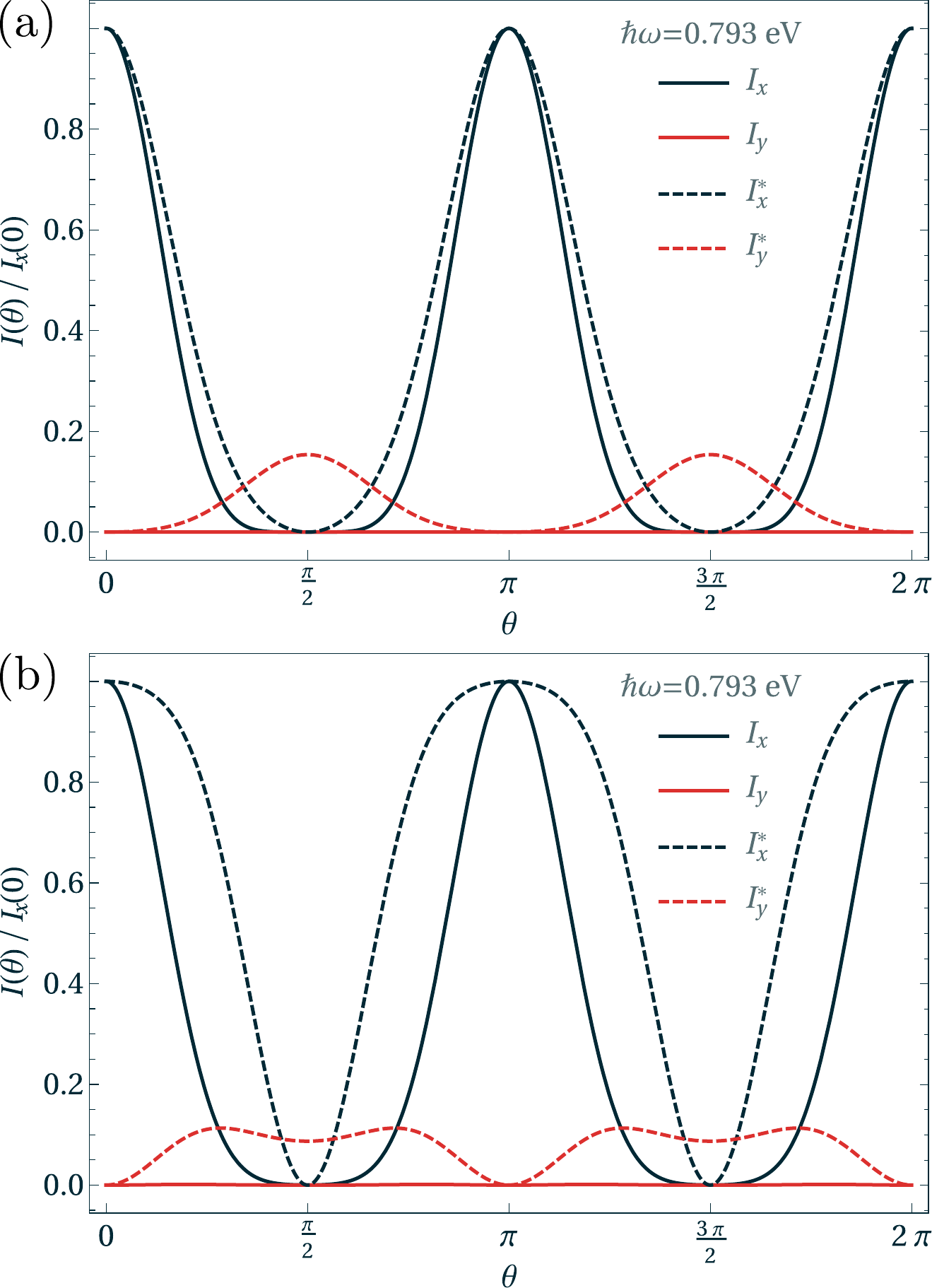}
\caption[THG intensities for bulk and monolayer]{\label{fig:5}%
Normalized THG intensities for bulk (a) and monolayer (b) bP at 
$\hbar\omega \teq 0.793 \, \mathrm{eV} \, \tsim 1560 \, \mathrm{nm}$.
Solid lines depict the THG intensity pattern using the nonlinear 
conductivities computed with \Eqref{eq:s3}.
Dashed lines depict the THG intensity pattern with increased response along 
$y$-axis as described in the main text.
Black and red lines represent the THG pattern along crystal directions, 
$\zeta= \{0,\pi/2 \}$.}
\end{figure}%

The analysis of the radiated THG signal, \Eqsref{eq:I3}, provides a tool to 
probe the nonlinear conductivity tensor. In \Fsref{fig:5}a and 
\ref{fig:5}b, we plot the normalized intensity patterns for bulk and monolayer 
bP.
Solid black (red) curves represent $I_x \, (I_y)$ intensities at incident 
photon energy $\hbar\omega=0.793$ eV, using results obtained from the 
evaluation of \Eqsref{eq:s3}.
The anisotropy of the system manifests itself clearly for both the bulk and 
monolayer bP, with the patterns dominated by the contribution of 
$I_x/I_0 \simeq |\bar\sigma_{11}|^2\cos^6\theta$.
To the best of our knowledge, experimental data on THG in bP is limited to bulk 
or several layer \cite{Rodrigues2016, Autere2017, Youngblood2017} and results 
for the intensity dependence on the polarization angle appear to be 
inconsistent, \eg pattern of total intensity presented by 
Ref.~\onlinecite{Youngblood2017} exhibits maxima along the crystal 
$x$-direction, whereas Refs.~\onlinecite{Rodrigues2016, Autere2017} shows 
maxima align with directions other than the primitive lattice directions, 
namely $\theta\sim \{ \pm \pi/6,\pm5\pi/6\}$.
Additionally, the pattern for $I_y$ in Ref.~\onlinecite{Youngblood2017} is not 
symmetric with respect to $y$-direction, \ie $\theta=\pm\pi/2$, hence not 
compatible with the THG radiated field by orthorhombic crystals, 
\Eqsref{eq:I3}.
Notwithstanding these differences between the experimental results, all 
indicate a much larger response along the $y$-direction ($\zeta-\pi/2$) than 
that predicted by our results.
Following the spirit of Ref.~\onlinecite{Low2014a}, we consider the effect of 
artificially increasing the matrix elements along the $y$ direction by a 
constant factor.
Such increase can make $I_y$ visible in the scale of \Fsref{fig:5}a and 
\ref{fig:5}b at $\zeta=\pi/2$ as depicted by dashed lines, where the 
$y$-direction matrix elements are increased by factors of $4.5$ and $3.25$, 
respectively.
Nonetheless, the new patterns remain inconsistent with reported experimental 
data, indicating that this discrepancy should stem from additional mechanisms. 
It is worth noting that recent results of photoluminescence in high quality 
samples \cite{Li2016a} have shown that the linear response along the 
$y$-direction is vanishingly small, indicating that the apparently higher 
response along the $y$ direction can be attributed to mechanisms other than the 
intrinsic response of the system, such as disorder.
In addition, the estimate of the magnitude of $\chi_\mathrm{eff}^{(3)}$ and 
its ratio with regards to graphene's $\chi_\mathrm{eff}^{(3)}$ remains an open 
question, as experimental reports indicate different results that span several 
orders of magnitude \cite{Rodrigues2016, Youngblood2017, Autere2017}.
Our results indicate that both bulk and monolayer THG conductivities at 
$\hbar\omega = 0.793$ eV ($\lambda\sim1560$ nm) have magnitudes 
$\sim20\sigma_3$, which corresponds to a nonlinear susceptibility 
$\chi_\mathrm{eff}^{(3)}\sim 10 \times 10^{-19} \, \mathrm{nm}^2/\mathrm{V}^2$, 
similar to recent reported results for bulk bP \cite{Youngblood2017, 
Autere2017}.
%

\section{Concluding remarks}

We studied THG in bP based on derivation of the nonlinear current density 
response, without the divergences that plague the direct evaluation of 
$j^{(3)}$ even when computed in the length gauge \cite{Aversa1995}.
We show that these divergences are spurious and can be removed by considering 
the effective tensor components, \ie physically relevant combinations of tensor 
elements, rather than the individual elements 
$\sigma_{\phi\lambda\beta\alpha}$. 
The resulting nonlinear conductivities, \Eqsref{eq:s3}, are free of 
divergences and can be applied directly to two-band systems in the independent 
particle approximation.
Using a non-orthogonal TB model to compute the energy dispersion and 
eigenstates of bP, we evaluate the low energy THG conductivity.
Results for bulk bP agree, at least qualitatively, with the experimental 
reports of THG in bulk or many layer samples bP \cite{Rodrigues2016, 
Youngblood2017, Autere2017}.

The present calculations ignore electron-electron interactions, which can play 
an important role in the optical response of a material, particularly for 
insulators with a large gap such as the hexagonal Boron Nitride (hBN), 
monolayers of TMDs, as well as mono- and few-layer bP.
It has been shown that, due to the large gap in hBN, excitonic binding plays a 
crucial role in SHG \cite{Pedersen2015} and nonlinear photocurrents 
\cite{Hipolito2016}. In both cases, the response onset is reduced significantly 
and most of of spectral weight is transfered to the features associated with 
the fundamental exciton.
First principles studies indicate that the linear response of single and 
few-layer bP exhibit similar behavior \cite{Tran2014, Tran2015}.
Therefore, our results for monolayer bP, computed within the framework of 
single particle approximation, should be considered as a qualitative 
description of the response, rather than quantitatively.
With respect to bulk bP, we expect excitonic effects to play a small role, 
since the exciton binding energy decreases with increasing number of layers 
\cite{Tran2014a, Tran2015}.
Experimental reports on photoluminescence \cite{Zhang2014} and extinction 
spectra \cite{Zhang2017} support the results of theoretical studies on the 
effects of electron-electron interations by showing that the excitonic 
resonances soften with increased number of layers.
Furthermore, the small gap of bulk bP facilitates doping with charge carriers, 
which in turn will suppress the electron-electron interactions even further.
This is supported by the smooth and step-like extinction spectra for bulk bP 
reported in Ref.~\onlinecite{Zhang2017} and also by the dielectric function of 
bulk bP computed from EELS data in Ref.~\onlinecite{Schuster2015}.
Based on these experimental reports and the above-mentioned arguments, we 
expect that the nonlinear response of bulk bP can be accurately characterized 
within the framework of the single particle approximation.

\acknowledgments 

The authors thank F. Bonabi and A. Taghizadeh for helpful discussions 
throughout this project. 
This work was supported by the QUSCOPE center sponsored by the Villum 
Foundation and 
TGP is financially supported by the CNG center under the Danish National 
Research Foundation, project DNRF103. 

\bigskip

\bibliography{bP-thg}

\end{document}